\documentclass[aps,prd,preprint]{revtex4-1}
\usepackage{amsmath}
\usepackage{amssymb}
\usepackage{enumitem}
\usepackage{color}
\usepackage{url}
\usepackage{xcolor}
\usepackage{graphicx}
\usepackage{verbatim}
\usepackage{tikz}
\usepackage{mathtools}
\usepackage{bbm}
\usepackage{accents}

\newcommand{\nc}{\newcommand}
\newcommand{\rnc}{\renewcommand}

\nc\bA{\mathbb{A}}
\nc\bB{\mathbb{B}}
\nc\bC{\mathbb{C}}
\nc\bD{\mathbb{D}}
\nc\bE{\mathbb{E}}
\nc\bF{\mathbb{F}}
\nc\bG{\mathbb{G}}
\nc\bH{\mathbb{H}}
\nc\bI{\mathbb{I}}
\nc{\bJ}{\mathbb{J}} 
\nc\bK{\mathbb{K}}
\nc\bL{\mathbb{L}}
\nc\bM{\mathbb{M}}
\nc\bN{\mathbb{N}}
\nc\bO{\mathbb{O}}
\nc\bP{\mathbb{P}}
\nc\bQ{\mathbb{Q}}
\nc\bR{\mathbb{R}}
\nc\bS{\mathbb{S}}
\nc\bT{\mathbb{T}}
\nc\bU{\mathbb{U}}
\nc\bV{\mathbb{V}}
\nc\bW{\mathbb{W}}
\nc\bY{\mathbb{Y}}
\nc\bX{\mathbb{X}}
\nc\bZ{\mathbb{Z}}
\nc\cA{\mathcal{A}}
\nc\cB{\mathcal{B}}
\nc\cC{\mathcal{C}}
\nc\cD{\mathcal{D}}
\nc\cE{\mathcal{E}}
\nc\cF{\mathcal{F}}
\nc\cG{\mathcal{G}}
\nc\cH{\mathcal{H}}
\nc\cI{\mathcal{I}}
\nc{\cJ}{\mathcal{J}} 
\nc\cK{\mathcal{K}}
\nc\cL{\mathcal{L}}
\nc\cM{\mathcal{M}}
\nc\cN{\mathcal{N}}
\nc\cO{\mathcal{O}}
\nc\cP{\mathcal{P}}
\nc\cQ{\mathcal{Q}}
\nc\cR{\mathcal{R}}
\nc\cS{\mathcal{S}}
\nc\cT{\mathcal{T}}
\nc\cU{\mathcal{U}}
\nc\cV{\mathcal{V}}
\nc\cW{\mathcal{W}}
\nc\cY{\mathcal{Y}}
\nc\cX{\mathcal{X}}
\nc\cZ{\mathcal{Z}}

\nc{\dmo}{\DeclareMathOperator}
\rnc{\Re}{\operatorname{Re}}
\rnc{\Im}{\operatorname{Im}}
\dmo{\Span}{Span}
\dmo{\rank}{rank}
\dmo{\End}{End}
\dmo{\Hom}{Hom}
\dmo{\Jac}{Jac}
\dmo{\Id}{Id}
\dmo{\Ann}{Ann}
\dmo{\Area}{Area}
\dmo{\CP}{\bC P^1}
\dmo{\Aut}{Aut}
\dmo{\Tr}{Tr}

\begin{document}
\title{Superrotations are Linkages}
\author{Ratindranath Akhoury}
\affiliation{Leinweber Center for Theoretical Physics, Randall Laboratory of Physics, Department of Physics, University of Michigan, Ann Arbor, MI 48109, USA}
\author{David Garfinkle}
\affiliation{Department of Physics, Oakland University, Rochester, MI 48309, USA}
\affiliation{Leinweber Center for Theoretical Physics, Randall Laboratory of Physics, Department of Physics, University of Michigan, Ann Arbor, MI 48109, USA}
\author{Arielle Schutz}
\affiliation{Leinweber Center for Theoretical Physics, Randall Laboratory of Physics, Department of Physics, University of Michigan, Ann Arbor, MI 48109, USA}

\date{\today}

\begin{abstract}
We show that superrotations can be described using the geometric conformal completion method of Penrose.  In particular, superrotation  charges and fluxes can be described and calculated using the linkage method of Geroch and Winicour.  

Whether superrotation charges are calculated using the coordinate based Bondi formalism or the geometric Penrose formalism, the fact that the superrotation blows up at a point makes the superrotation charge formally ill defined.  Nonetheless, we show that it can be made well defined through a regularization procedure devised by Flanagan and Nichols.
\end{abstract}

\maketitle
\section{Introduction}

The Bondi-Sachs metric is a way to describe spacetimes that are asymptotically flat at null infinity. The asymptotic symmetries of this metric at null infinity form a group known as the BMS group generated by the vector field $\xi^a$. To each vector $\xi^a$ and each cross section of the boundary at null infinity is associated a charge $Q$.

Here, we focus on an expansion of the BMS group suggested by Banks \cite{Banks_2003}, Barnich and Troessart \cite{Barnich_2010b,Barnich_2012} and Hawking, Perry and Strominger \cite{Hawking_2016}. This expansion consists of \textit{supertranslations} and \textit{superrotations}, which are functions at the conformal 2-sphere at infinity. The supertranslations are well behaved; however, for superrotations these functions introduce some difficulties because of their singularities. For a review of superrotations, see \cite{Strominger_2017a} and \cite{Pasterski_2019}.

An alternative way to describe asymptotically flat spacetimes is the conformal completion technique of Penrose \cite{Penrose:1962ij}, an elegant geometric technique in which one introduces an unphysical metric that has a boundary at null infinity and that is conformally related to the physical metric.  All equations are formulated in the unphysical spacetime, and in particular all quantities to be evaluated ``in the limit as one goes to null infinity'' are simply evaluated on the boundary.

A number of recent works have revisited the charges defined by BMS symmetries. Nevertheless there are divergences in the calculation of such charges. Compère, Fiorucci, and Ruzziconi\cite{Compere_2020} propose modified boundary conditions to regularize such divergences. Additionally, in the work of \cite{Freidel_2021}, phase space renormalization is used for the same purpose.

As described by Fiorucci, these divergences may also be discussed in the view of leaky phase spaces \cite{Fiorucci_2021}. Furthermore, Rignon-Bret and Speziale \cite{RignonBret_2024,RignonBret_2025} describe this problem in the framework of algebraic topology as field-dependent 2-cocycles.

There are also conceptual questions concerning the covariance of asymptotic charges. Recent work by Chen, Wald, Yau et. al. \cite{Chen:2022fbu} has emphasized the importance of ``cross-section continuity" as a criterion for angular momentum at null infinity. In particular, a covariant charge $Q$ satisfying a relation of the form $dQ = F$, where $F$ is a covariant flux, will obey cross-section continuity. At the same time, it has been argued that there is no local and covariant construction of the symplectic current at null infinity \cite{Flanagan_2020}, suggesting an obstruction to defining fully covariant asymptotic charges. It is therefore natural to ask whether superrotation charges can be defined in a manifestly covariant manner consistent with cross-section continuity.

In this paper we propose an alternative way to regularize such divergences, revisiting the manifestly-covariant approach of linkages described by Geroch and Winicour. In particular, as shown by Geroch and Winicour \cite{Geroch_1981}, the charges and fluxes of the BMS group can be evaluated as an integral on the boundary, provided that one adds a simple condition on the divergence of the BMS symmetry vector.

One natural question then is whether the method of Geroch and Winicour can also be applied to superrotations.  We will show that it can.  In section \ref{bondi} we review the usual Bondi coordinate description of asymptotically flat spacetimes, the BMS group and its generalization to superrotations.  In section \ref{conformal} we cover the Penrose method and its application by Geroch and Winicour, and we show that this method also works for superrotations.  In section \ref{regularize} we cover the regularization procedure that renders superrotation charges finite and well defined. Conclusions are given in section \ref{conclusion}.

\section{Bondi-Sachs metric}
\subsection{Asymptotically flat metrics}
\label{bondi}
Here, we describe the formalism of the metric introduced by Bondi and developed by Sachs \cite{Bondi_1960,Bondi_1962,Sachs_1961,Sachs_1962a,Sachs_1962b}, which is a useful way to express asymptotically flat metrics. Here we will focus on $d=4$.  The Bondi method uses a radial coordinate $r$, two angular coordinates $(\theta^1,\theta^2)$, and a null coordinate $u$ which can be conceptualized as something like $t-r$.  Null infinity is the limit as $r\to \infty$ at finite $(u,\theta^1,\theta^2)$.  So in a neighborhood of null infinity one expands all components of the metric in powers of $r^{-1}$.  For ease of comparison with the Penrose method, we will introduce the coordinate $\Omega \equiv {r^{-1}}$ and expand in powers of $\Omega$.  So in coordinates  $(u,\Omega,\theta^1,\theta^2)$, the line element takes the form
\begin{equation}
    d {s^2} = {\Omega ^{-2}} \left [ -U\Omega^2e^{2\beta}du^2+2e^{2\beta}dud\Omega
    +\gamma_{AB}(d\theta^A-\cU^Adu)(d\theta^B-\cU^Bdu) \right ] 
\end{equation}
where capital Roman letters refer to the angular indices 1,2. Because null infinity (hereafter denoted by $\cI^+$) has the topology $S^2\times\mathbb{R}$, it is evident that the angular indices $\theta^A$ refer to the 2-sphere $S^2$, and $\mathbb{R}$ is associated with the value of $u$. We write the usual round metric on $S^2$ as $h_{AB}$, which raises and lowers the capital Roman indices and which is associated with the covariant derivative $D_A$.

Now, we expand the metric functions \cite{Barnich_2011}. Asymptotic flatness ensures the order of $\Omega$ at which the expansions of $(U,\beta,\cU^A,\gamma_{AB})$ start.
\begin{align}
    \beta &= \beta_0\Omega+\beta_1\Omega^2+\beta_2\Omega^3+\cO(\Omega^4),\\
    U &= 1-2m\Omega-2\cM\Omega^2+\cO(\Omega^3),\\
    \gamma_{AB} &= h_{AB} + C_{AB}\Omega+D_{AB}\Omega^2+E_{AB}\Omega^3+\cO(\Omega^4),\\
    \cU^A &= U^A\Omega^3+\Omega^3\left[-\frac{2}{3}N^A+\frac{1}{16}D^A(C_{BC}C^{BC})\right.\nonumber\\
    &+\left.\frac{1}{2}C^{AB}D^C C_{BC}\right]+\cO(\Omega^4).
\end{align}
We note the presence of important radiative data in the expansion: the Bondi mass aspect $m(u,\theta^A)$, the angular momentum aspect $N^A(u,\theta^A)$, and the tensor $C_{AB}(u,\theta^A)$. Here, $C_{AB}$ is symmetric and its derivative is the Bondi news tensor is $N_{AB} = \partial_u C_{AB}$. We also must impose the gauge conditions
\begin{align}
    \tilde{g}_{\Omega\Omega} &= \tilde{g}_{\Omega A} = 0,\nonumber\\
    \partial_\Omega\det\gamma_{AB} &= 0 \label{gauge}
\end{align}
to get the tracelessness of $C_{AB}$, as well as Einstein's equations, to eventually write the components of the metric $g_{ab}$ in terms of $(u,\Omega,\theta^A)$ from our expansion earlier \cite{Flanagan_2017}:
\begin{align}
    \tilde{g}_{uu} &= \Omega^{-2}\left[-\Omega^2+ 2M\Omega^3 + 2\cM\Omega^4 + C_{AB}C^{AB}\Omega^4/16\right],\nonumber\\
    \tilde{g}_{u\Omega} &= \Omega^{-2}\left[1-C_{AB}C^{AB}\Omega^2/16,\right]\nonumber\\
    \tilde{g}_{\Omega\Omega} &= \tilde{g}_{\Omega A} = 0,\\
    \tilde{g}_{AB} &= \Omega^{-2}\left[h_{AB}+\Omega C_{AB}+\Omega^2 D_{AB}+\Omega^3 E_{AB} + \Omega^4 F_{AB}\right],\nonumber\\
    \tilde{g}_{uA} &= \Omega^{-2}\left[-\Omega^2 U_A-\Omega^3\left(C_{AB}U^B-\frac{2}{3}N_A+\frac{1}{16}D_A(C_{BC}C^{BC})\right)+ \cO(\Omega^2)\right].\nonumber
\end{align}
Here,
\begin{equation}
    U_A \equiv -D^BC_{AB}/2.
\end{equation}
\subsection{BMS symmetry group}
The group of asymptotic symmetries of an asymptotically flat metric is called the Bondi-Metzner-Sachs (BMS) group. This was described by Bondi, Van der Burg, and Metzner \cite{Bondi_1962} and expanded upon by Sachs \cite{Sachs_1962b}. This is often referred to in the literature as $\mathfrak{bms}_d$. In this section, we will introduce this group and describe its extension to the superrotations.

Let $\xi^a$ be the vector field that generates $\mathfrak{bms}_3$. That is, one imposes that ${\cal L}_\xi$ of the Bondi metric is zero to the appropriate order in $\Omega$.  The result is that on null infinity $\xi^a$ takes the form
\begin{align}
    \xi^u &= \alpha(\theta^A)+\frac{1}{2}uD_AY^A(\theta^B),\\
    \xi^A &= Y^A(\theta^B),
\end{align}
where $Y^A$ is a conformal Killing vector (CKV) of the two-sphere.  There are six smooth conformal Killing vectors of the two-sphere, which correspond to the Lorentz group.  The special cases where $\alpha$ is a linear combination of $\ell=0$ and $\ell=1$ spherical harmonics are called translations and correspond to the usual four-parameter translation group.  However, because $\alpha$ can be any function of the angles, the usual ten-parameter Poincaré group is enlarged to the infinite-dimensional BMS group with the extra symmetries called supertranslations. 

Then, $\xi^a$ can be extended into the interior of the manifold to produce approximate Killing vectors inside the manifold by ensuring that the falloff conditions
\begin{equation}
    \cL_\xi\tilde{g}_{ur} = \cO(r^{-2}),\cL_\xi\tilde{g}_{uz} = \cO(1),\cL_\xi\tilde{g}_{zz} = \cO(r),\cL_\xi\tilde{g}_{uu} = \cO(r^{-1})
\end{equation}
are met. We thus obtain obtain\cite{Sachs_1962b}\cite{Barnich_2011}\cite{Flanagan_2017}
\begin{align}
    \xi^a &= f\partial_u + \left[Y^A-\Omega D^Af+\frac{1}{2}\Omega^2C^{AB}D_Bf+\cO(\Omega^3)\right]\partial_A\nonumber\\
    &-\left[\frac{\Omega}{2}D_AY^A-\frac{\Omega^2}{2}D^2f-\frac{\Omega^3}{2}\Omega U^A D_Af\right.\\
    &+\left.\frac{1}{4}\Omega^3 D_A(D_BfC^{AB})+\cO(\Omega^4)\right]\partial_\Omega\nonumber,
\end{align}
where
\begin{equation}
    f(u,\theta^A) \equiv \alpha(\theta^A) + \frac{1}{2}uD_BY^B(\theta^A).
\end{equation}

The proposal of Banks and of Barnich and Troessart is essentially to further enlarge the BMS group by allowing conformal Killing vectors of the two-sphere that are not smooth at the north pole.  In analogy with the supertranslations that enlarged the Poincare group to form the BMS group, these extra symmetries are called superrotations.  

To see how this works, note the following: (1) two-dimensional spaces are locally conformally flat, and (2) a conformal Killing vector is a conformal invariant.  Thus if one could get rid of the ``locally'' in locally conformally flat, a two dimensional space would have as many CKVs as two dimensional flat space.  Now specialize to the case of the two-sphere without the north pole and use the usual $(\theta,\phi)$ coordinates so that the line element takes the form
\begin{equation}
d {s^2} = d {\theta ^2} + {\sin ^2}\theta d {\phi^2}.    
\end{equation}
Define $x \equiv \cot (\theta/2)\cos \phi$ and $y \equiv \cot (\theta/2) \sin \phi$.  Then we obtain
\begin{equation}   
d {\theta ^2} + {\sin ^2}\theta d {\phi^2}    = {\frac 4 {1+x^2+y^2}} \left (  d {x^2} +d {y^2}   \right ).
\end{equation}
It is helpful to introduce the complex coordinate $z=x+iy$ because the conformal Killing vectors of two dimensional flat space are those whose $z$ component is an analytic function of $z$.  In terms of the origonal two-sphere coordinates we have
$z = \cot(\theta/2)e^{i\phi}$. Putting our expressions in terms of $z$  the two-sphere metric can be written as
\begin{align}
    \begin{pmatrix}
        0 & \frac{2}{(1+z\bar{z})^2}\\
        \frac{2}{(1+z\bar{z})^2} & 0
    \end{pmatrix}.
\end{align}
and the CKV equations are
\begin{equation}
    \partial_z Y^{\bar{z}} = \partial_{\bar{z}} Y^z = 0.
\end{equation}
Thus, the generators of the solutions can be simply written in terms of the basis
\begin{equation}
    \{l_m \equiv z^m\partial_z, \bar{l}_m \equiv \bar{z}^n\partial_{\bar{z}}\},
\end{equation}
for $m,n\in \mathbb{Z}$. Here, the smooth, finite solutions are precisely the Lorentz group, and they are represented by the basis $\{l_{-1},l_0,l_1,\bar{l}_{-1},\bar{l}_0,\bar{l}_1\}$. This was noted by Sachs \cite{Sachs_1962b}.

When we expand the scope of solutions under consideration to any solution of the CKV equations, we obtain the \textit{superrotations}. Since any meromorphic function can be expanded in a Laurent series, this is equivalent to considering general values of $m\in\mathbb{Z}$. In other words, the superrotations are the full, infinite-dimensional algebra of conformal transformations of $S^2$.

For a purely rotational $\xi$ with $\alpha(\theta^A) \equiv 0$,
\begin{align}
    \xi &= \frac{1}{2}uD_AY^A\partial_u +\left[\frac{\Omega}{2}D_AY^A-\frac{u\Omega^2}{4}D^2D_AY^A +\frac{u\Omega^3}{8}U^AD_AD_CY^C\right.\nonumber\\
    &+\left.\frac{u\Omega^3}{8}D_A((D_BD_CY^C)C^{AB})\right]\partial_\Omega + Y^A\partial_A \label{xi}\\
    &+\left[-\frac{u\Omega}{2}D^AD_BY^B+\frac{u\Omega^2}{4}C^{AB}D_BD_CY^C\right]\partial_A\nonumber.
\end{align}

Our motivation for considering these solutions was previously proposed by Banks \cite{Banks_2003} and Barnich and Troessart \cite{Barnich_2010b}\cite{Barnich_2012}. Though these solutions are not globally well defined, they have an important relationship to the Witt algebra and its extension, the Virasoro algebra, namely \cite{Capone_2018}
\begin{equation}
    \mathfrak{bms}_4 = (\mathfrak{witt}\times\mathfrak{witt})\oplus(\mathfrak{witt}\times\mathfrak{witt}).
\end{equation}
What is the relevance of these superrotations? Strominger and Zhibodoev \cite{Strominger_2017} provided a physical interpretation of superrotations: spacetimes before and after a string decay differ by a finite superrotation.

Campiglia and Laddha \cite{Campiglia_2014}\cite{Campiglia_2015} proposed defining superrotations as $\text{Diff}(S^2)$ instead. The advantage of this approach is that first, it can easily be extended to $d$ dimensions as $\text{Diff}(S^{d-2})$. Moreover, they argued that the Ward identities associated with the generators of this group are equivalent to the subleading soft graviton theorem showed by Cachazo and Strominger \cite{Cachazo_2014}. However, these symmetries do not arise as infinitesimal symmetries of the Bondi-Sachs metric unless the boundary conditions are also changed \cite{Campiglia_2014}\cite{Campiglia_2015}\cite{Compere_2018}\cite{Chandrasekaran_2018}. In our article, will consider the group proposed by Barnich and Troessart, the group $\mathfrak{bms}_4$ with basis $\{l_m,\bar{l}_m\}$ for $m\in\mathbb{Z}$.

\section{The Penrose conformal method}
\label{conformal}
A different approach to asymptotic flatness from the Bondi method is the Penrose conformal method.  Though these two approaches at first look very different, they turn out to be equivalent.  In the Penrose approach the physical spacetime is denoted  $(\tilde{M}_{ab},\tilde{g}_{ab})$, and it is said to be asymptotically flat provided that there is an unphysical spacetime $(M,g_{ab})$ with boundary $I$ such that the following conditions are satisfied: 
\begin{enumerate}[label={\arabic*.}]
    \item On $M$,
\begin{enumerate}
    \item $\tilde{M}$ is diffeomorphic to the interior $M - I$ of $M$,
    \item there is a smooth function $\Omega$ on $M$ such that $g_{ab} = \Omega^2\tilde{g}_{ab}$ everywhere on $M$.
\end{enumerate}
    \item At the boundary $I$,
\begin{enumerate}
    \item $\Omega = 0$,
    \item the normal vector $\nabla_a\Omega\neq 0$, and
    \item $\nabla_a\Omega\nabla^a\Omega = 0$.
\end{enumerate}
\end{enumerate}
Condition 1(a) means that we can compactify the manifold $\tilde{M}$ into one defined at a boundary, $M$. This, along with Condition 1(b) means that the conformal factor $\Omega$ rescales the \textit{physical metric} $\tilde{g}_{ab}$ into the \textit{unphysical metric} $g_{ab}$.

Condition 2(a) means that $\Omega$ vanishes at the boundary. In fact, paired with the previous conditions, it means that $\Omega = 0$ exactly at the boundary. Along with 2(b), this fixes the scaling of $\Omega$ as $\Omega\tilde 1/r$. Along with 2(b) and 2(c), this means that the boundary is a null surface, since its only normal vectors are null vectors. As a consequence of asymptotic flatness, the boundary must have the topology $S^2\times\mathbb{R}$. We identify the boundary $I$ with null infinity, in particular, future null infinity $\cI^+$. In this paper, we shall set $\Omega\equiv 1/r$ and use $\Omega$ instead of $r$ for the radial coordinate.

The reason for this backwards-looking notation (tilde for physical quantities and no tilde for unphysical quantities) is that to the extent possible one wants to perform all calculation in the unphysical spacetime and obtain ``limits at null infinity'' by simply evaluating quantities at $\cI^+$.

If the physical spacetime had a Killing vector, it would satisfy 
${\cal L}_\xi {{\tilde g}_{ab}} =0$.  However, BMS symmetries are supposed to be close enough to flat spacetime Killing vectors in the limit as null infinity is approached.  So for a BMS generator, the definition is a vector field $\xi ^a$ such that ${\Omega ^2}{\cal L}_\xi {{\tilde g}_{ab}}$ is smooth and vanishes on $\cI^+$.  It then follows that there is a smooth scalar $K$ and tensor $X_{ab}$ such that ${\xi ^a}{\nabla _a}\Omega = \Omega K$ and 
\begin{equation}
{\nabla _{(a}}{\xi _{b)}} = K {g_{ab}} + \Omega {X_{ab}}
\label{GerochBMS}
\end{equation}

To get a feel for the equivalence of the two approaches to asymptotic flatness, it is instructive to verify that the components of $X_{ab}$ are indeed smooth for a metric of the Bondi form and a $\xi^a$ satisfying the Bondi description of a BMS generator.  We use the usual $(\theta,\phi)$ coordinates for the two-sphere and introduce the notation $\Theta \equiv {\xi^\theta} ={Y^\theta}$ and $\Phi \equiv {\xi^\phi}={Y^\phi}$.  
Then the terms in $X_{ab}$ that go like $\Omega^{-1}$  are zero, except for the terms
\begin{align}
    X_{\theta\theta} &= {\Omega^{-1}}\frac{1}{2}(-D_AY^A+2\partial_\theta \Theta),\nonumber\\
X_{\theta\phi} &= {\Omega ^{-1}} X_{\phi\theta} = \frac{1}{2}(\partial_\phi \Theta + \sin^2\theta\partial_\theta \Phi),\\
X_{\phi\phi} &= {\Omega ^{-1}} \frac{1}{2}(-D_AY^A\sin^2\theta+2\sin\theta\cos\theta \Theta + 2\sin^2\theta \partial_\phi \Phi)\nonumber.
\end{align}
Here ${D_A}{Y^A}={\partial _\theta}\Theta + \Theta \cot \theta + {\partial _\phi} \Phi$.
However, the equations for a two-sphere conformal Killing vector are
\begin{eqnarray}
{\partial _\theta}\Phi &=& {-\frac {1} {{\sin^2} \theta}} {\partial _\phi}\Theta
\nonumber
\\
{\partial _\phi}\Phi &=& {\partial _\theta}\Theta - \cot {\theta} \Theta  \end{eqnarray}
After imposing our CKV relations we see that the terms in $X_{ab}$ that threaten to be singular all have coefficients of zero and thus that $X_{ab}$ exists and is well defined. This is, of course, expected as $\xi^a$ must be a generator of BMS symmetry.

\section{Charges, fluxes, and linkages}
\subsection{Charges}
\label{regularize}
If $\xi^a$ were a Killing vector of the physical spacetime, then the amount of its charge contained within a two-surface $S$ would be given by the Komar formula\cite{Komar_1958}
\begin{equation}
Q = {\int _S} {{\tilde \epsilon}_{abmn}} {{\tilde {\nabla}}^m}{\xi^n}  
\end{equation}
The linkage strategy of Geroch and Winicour\cite{Geroch_1981} is to use the Komar formula for BMS symmetries, but now to express everything in terms of the unphysical metric and to have the surface $S$ be a cross-section of $\cI^+$.  
Similar strategies were later used by Wald and Zoupas \cite{Wald_2000} and by Crnkovic and Witten \cite{Crnkovic_1986}. The charge is
\begin{equation}
    Q = \int_S\epsilon_{abmn}\nabla^m(\Omega^{-2}\xi^n),\label{asy_charge}
\end{equation}
where $S$ is a cross section of $\cI^+$, parametrized by the value of $u$.

One difficulty in evaluating the integral for $Q$ is the presence of the inverse powers of $\Omega$, which diverge at the boundary. 
This difficulty is addressed by Geroch and Winicour as follows: they first define $Q_0$ to be the integral over a sphere $S_0$ in the bulk rather than at infinity.  This integral is finite since $S_0$ is not at infinity.  They then note that the difference between $Q$ and $Q_0$ is 
the integral over a three-dimensional surface connecting $S$ to $S_0$
of $d({\epsilon _{abmn}} {\nabla ^m}({\Omega ^{-2}} {\xi^n}))$. 
They then show that despite the presence of the $\Omega ^{-2}$ in this expression the quantity $d({\epsilon _{abmn}} {\nabla ^m}({\Omega ^{-2}} {\xi^n}))$ is finite at $\cI^+$.  

This strategy could also be used for superrotations provided that one can show also in the superrotation case that $d({\epsilon _{abmn}} {\nabla ^m}({\Omega ^{-2}} {\xi^n}))$ is finite. Since the superrotations themselves are infinite at the north pole (and at the south pole for vector fields using terms proportional to negative powers of $z$) we must exclude a neighborhood of the north pole (and the south pole) and then demonstrate that there are no terms proportional to negative powers of $\Omega$.  This demonstration is given in the appendix.

Another issue addressed by Geroch and Winicour is that of gauge: for an asymptotically flat spacetime the conformal factor used in its completion is not unique.  Thus any proposed formula for the charge should be invariant under change of conformal factor.  Furthermore, the condition of eqn. (\ref{GerochBMS}) does not uniquely determine how $\xi^a$ extends into the bulk.  Thus any proposed formula for the charge should place some restrictions on how $\xi$ extends into the bulk in order that it be unique. 

The result of Geroch and Winicour is that both these gauge issues can be addressed by simply imposing the condition $X=0$ where 
$X\equiv g^{ab}X_{ab} $ is the trace of the tensor $X_{ab}$.  This condition makes the charge formula independent of any remaining gauge.  

Note that the usual coordinate treatment of superrotations gives an expansion for $\xi^a$ that gives some information on how $\xi^a$ is to be extended into the bulk.  This expansion turns out to have as a consequence that the Geroch-Winicour gauge condition $X=0$ is satisfied.  Explicitly, the expansion implies that the value of $X$ at $\cI^+$ is
\begin{align}
    X &= 1/2(-2a(\cot\theta \Theta + \partial_\phi \Phi - \partial_\theta \Theta)+2c(\csc^2\theta\partial_\phi \Theta+\partial_\theta \Phi)\nonumber\\
&-u\csc^2\theta(\cot\theta\partial^2_\phi \Theta + \partial^3_\phi \Phi + \partial_\theta\partial^2_\phi \Theta)\nonumber\\
&-u\cot\theta(-\csc^2\theta \Theta +\cot\theta\partial_\theta \Theta +\partial_\theta\partial_\phi \Phi + \partial^2_\theta \Theta)\nonumber\\
&-u(2\cot\theta\csc^2\theta \Theta - 2\csc^2\theta\partial_\theta \Theta + \cot\theta\partial^2_\theta \Theta + \partial^2_\theta\partial_\phi \Phi + \partial^3_\theta \Theta)).
\end{align}
Imposing the CKV relations, we get that $X = 0$. [*So therefore this condition is equivalent to the falloff condition of $\xi$ which is also expressed in Geroch and Winicour \cite{Geroch_1981} and Flanagan and Nichols \cite{Flanagan_2017}.]

Finally, we consider the fact that since superrotations are singular at the north pole (or the south pole or both) the integrand for the charge is singular, and thus the integral for the charge is an improper integral.  An improper integral can nonetheless be well defined provided that the answer does not depend on the particular limiting procedure used include the whole region of integration.  However, this is not the case for superrotations: as we will see finiteness depends on a delicate cancellation between positive and negative terms.  Nonetheless, if one particular limiting procedure produces a finite result, then this way of taking the limit can be used as a regularization procedure to make a formally ill defined quantity well defined.

Such a regularization procedure was devised by Flanagan and Nichols \cite{Flanagan_2017} though they stated it as a demonstration of finiteness, rather than a regularization procedure.  We now recapitulate the argument of \cite{Flanagan_2017} though stated more carefully as a regularization procedure.  

The integral is
\begin{align}
    Q &= \frac{1}{8\pi G}\int d^2\Omega \left[MuD_AY^A-\frac{1}{8}(D_AY^A)(C_{BC}C^{BC})+\frac{1}{2}u U^BD_B(D_AY^A)\right.\nonumber\\
&+3\left(\frac{2}{3}N_A-\frac{1}{16}D_A(C_{BC}C^{BC})-\frac{1}{2}{C_A}^B D^C C_{BC}\right)Y^A\\
&+\left.{C^A}_B U_A Y^B+\frac{1}{16}u(D_BY^B)(C^{CD}\partial_u C_{CD})\right]\nonumber.
\end{align}
We now exclude the region $\theta < \epsilon$ and $\theta > \pi-\epsilon$ and only take the limit as $\epsilon \to 0$ at the end.  Furthermore we expand the superrotation vector field in powers of $z$ and $\bar z$ and we expand all smooth quantities in scalar, vector, and tensor spherical harmonics.

Now, let's consider the singularities in the integral. Every term in the integral that does not depend on $Y^A$ should be nonsingular. Where $Y^A = z^n\partial_z$, the $Y^A$ terms are of the form
\begin{enumerate}[]
    \item $D_AY^A = (-1+n-\cos\theta)z^{n-1}$
    \item $D_B(D_AY^A)$, in which case we have $D_z(D_AY^A) = \frac{1}{4}(3-8n+4n^2-4(-1+n)\cos\theta+\cos(2\theta))z^{n-2}$ or
    $D_{\bar{z}} (D_A Y^A) = =-2(\sin(\theta/2))^4z^n$
    \item $Y^A = 0 \text{ or } z^n$
    \item $Y^B Y^A = 0 \text{ or } z^{2n}$
    \item $Y^A (D_BY^B) = 0 \text{ or } (-1+n+\cos\theta)\cot^n(\theta/2)\exp(\pm i\phi n)$
\end{enumerate}
When we have $Y^A = \bar{z}^n\partial_{\bar{z}}$ instead, then we just get the conjugate of these terms. The other terms in the integral are not dependent on $Y^A$ and must be nonsingular. The only possible terms that could lead to divergences are going to have the terms $\cot^{|n|}(\theta/2)e^{in\phi}$. We can expand all nonsingular terms in terms of spherical harmonics $Y^m_\ell = e^{im\phi}P_\ell(\cos\theta)$. Therefore we must integrate something like 
\begin{equation}
    \sum_{m,n = -\infty}^\infty \int d^2\Omega e^{im\phi}P^m_\ell(\cos\theta)f(\theta)e^{in\phi}(\cot(\theta/2))^{|n|}.
\end{equation}
Doing the $\phi$ integral gets us zero unless $m+n=0$, so then we are doing the integral
\begin{equation}
    2\pi \sum_{n=-\infty}^\infty \int d\theta \sin\theta P^n_\ell(\cos\theta) \cot(\theta/2)^{|n|}.
\end{equation}
Note that $P^n_\ell(\cos\theta) = f^{|n|}_\ell(\cos\theta)(\sin\theta)^{|n|}$, where $f$ is some polynomial. Therefore, we obtain
\begin{equation}
    4\pi \sum_{n=0}^\infty \int d\theta \sin\theta f^n_\ell(\cos\theta)(\sin\theta)^n\cot(\theta/2)^n = 4\pi\sum_{n=0}^\infty \int d\theta\sin\theta f^n_\ell(\cos\theta)(\cos\theta+1)^n,
\end{equation}
which is nonsingular. Therefore the singularities in $Y^A$ are not a problem.
\subsection{Asymptotic fluxes}

We will now address the question of whether the asymptotic integral of equation (\ref{asy_charge}) possesses an \textit{asymptotic flux}, that is, whether or not there exists a scalar function $F$ such that $Q_{S_1}-Q_{S_2} = \int_{S'}F$, where $S'$ is the region between surfaces $S_1$ and $S_2$. Geroch and Winicour\cite{Geroch_1981} address this question to show that there indeed exists a nonzero flux. With the gauge condition $X = 0$, this can be written as
\begin{equation}
    F = -\nabla^a\nabla^bX_{ab}+3\nabla^aX_a + (3/4)\nabla^2X+(1/24)RX.
\end{equation}
We note, as Geroch and Winicour do, this is locally defined, gauge invariant, and independent of cross sectional choice.  Though Geroch and Winicour obtained this formula for the case of BMS vector fields $\xi$, the formula generalizes to the case where $\xi$ is a superrotation.

Wald and Zoupas\cite{Wald_2000} express the differential flux, that is, the integrand of the expression $\int_{S'}F$, in terms of the simple expression
\begin{equation}
    (F_\xi)_{abc} = \Theta(\phi,\cL_\xi\phi)_{abc},
    \label{WZ}
\end{equation}
where $\Theta$ is the symplectic potential of the tensor fields of the theory, which are collectively denoted $\phi$. For general relativity with asymptotic flatness, Wald and Zoupas give an expression of this quantity in terms of the Bondi news tensor $N_{ab}$ and the perturbation $g_{ab}-g_{ab,\text{flat}}$ relative to flat spacetime in unphysical space. This quantity is
\begin{equation}
    \Theta_{cde} = -\frac{1}{32\pi}\Omega^{-1}N_{ab}\tau^{ab}{}^{(3)}\bar{\epsilon}_{cde},
\end{equation}
where the tensor ${}^{(3)}\bar{\epsilon}_{abc}$ is the pullback of the tensor ${}^{(3)}\epsilon_{abc}$ to $\cI^+$, and the latter is defined through the relation
\begin{equation}
    \epsilon_{abcd} = 4{}^{(3)}\epsilon_{[abc}n_{d]}.
\end{equation}
The difficult part of the Wald and Zoupas computation is finding the appropriate symplectic potential $\Theta$ for asymptotically flat general relativity.  Once one has the symplectic potential, the flux for the vector field $\xi$ is simply a matter of using $\xi$ in equation (\ref{WZ}).  This expression works as well when $\xi$ is a superrotation as it does in the original case where $\xi$ is a BMS symmetry.

\section{Conclusions}
\label{conclusion}
We have shown that the linkage method of Geroch and Winicour can be extended from the BMS group to the superrotations.  This provides a simple geometrical expression for the superrotation charge and flux.  Though this expression is formally ill defined, it can be made well defined by applying the regularization procedure of Flanagan and Nichols.

We have also shown that the superrotation charge is covariant. Chen, Wald, Yau et al. \cite{Chen:2022fbu} propose a notion of angular momentum based on “cross-section continuity,” which is satisfied by any covariant charge $Q$ for which $dQ = F$ for some covariant flux $F$. In particular, the CWY charge \cite{Chen:2013kza,Chen:2013lza} satisfies this condition, and coincides with the superrotation charge considered here. At the same time, it has been argued that no local and covariant definition of the symplectic current at null infinity exists \cite{Flanagan_2020}, suggesting a potential obstruction to constructing fully covariant asymptotic charges. By formulating superrotations within the manifestly covariant linkage framework, we demonstrate that the associated charge is indeed covariant and compatible with cross-section continuity.

\acknowledgments
We would like to thank Bob Wald, David Nichols, and Gautam Satishchandran for helpful discussions. The work of RA and AS is supported in part by the US Department of Energy, Office of Science award number DE-SC0025510. The work of DG was supported by NSF Grant PHY-2102914 to Oakland University.
\appendix
\section*{Appendix}
\section{Showing the linkage integrand is finite}\label{linkage}
We wish to show that the integrand $(dQ_L)_a$ of the linkage integral does not pose any problems related to the inverse powers of $\Omega$. Specifically, we show that the terms proportional to inverse powers of $\Omega$ vanish.

We may proceed by calculating $(dQ_L)_a$ and writing the result in various powers of $\Omega$:
\begin{equation}
    (dQ_L)_a \equiv \sum \Omega^n (R_n)_a.
\end{equation}
The only problematic terms here are when $n= -1,-2$. Here, we use the spherical coordinates $(\theta^1,\theta^2) = (\theta,\phi)$, with the metric $(1,\sin^2\theta)$. First of all, the tracelessness of $C_{AB}$ allows us to parametrize $C_{AB}$ in terms of  $a \equiv C_{\theta\theta} = -\csc^2\theta C_{\phi\phi}$ and $c \equiv C_{\theta\phi} = C_{\phi\theta}$.

In this coordinate system, let us define $\Theta \equiv Y^\theta$ and $\Phi \equiv Y^\phi$. the CKVs are
\begin{align}
    f_1\equiv -\partial_\theta \Theta +\cot\theta \Theta + \partial_\phi \Phi &= 0,\nonumber\\
    f_2\equiv \partial_\phi \Theta + \sin^2\theta\partial_\theta \Phi &= 0.
\end{align}

For $(R_{-2})_a$, we have $(R_{-2})_u = 0$ automatically and
\begin{align}
    (R_{-2})_\Omega &= \frac{1}{2}(-af_1 + cf_2) = 0,\\
    (R_{-2})_\theta &= \left(\cot\theta+\frac{1}{2}\partial_\theta\right)f_1-\frac{1}{2}\partial_\phi f_2 = 0,\\
    (R_{-2})_\phi &= -\frac{1}{2}\partial_\phi f_1+\left(-\frac{3}{2}\sin\theta\cos\theta-\frac{1}{2}\sin^2\theta\partial_\theta\right)f_2 = 0.
\end{align}

Now, let's consider $(R_{-1})_a$. First,
\begin{align}
    (R_{-1})_u &= \left[-\frac{1}{2}\partial_u a+\frac{1}{2}\partial^2_\theta-\frac{1}{2}\csc^2\theta\partial^2_\phi+\frac{3}{2}\cot\theta\partial_\theta-1\right]f_1\nonumber\\
&+\left[\frac{1}{2}\partial_u c-\partial_\theta\partial_\phi-2\cot\theta\partial_\phi\right]f_2 = 0.
\end{align}
The component $(R_{-1})_\Omega$ is automatically zero. Next,
\begin{align}
    (R_{-1})_\theta &= \left[\left(\frac{1}{2}\partial_\theta+\cot\theta\right)a-c\csc^2\theta \partial_\phi-\frac{1}{2}\csc^2\theta \partial_\theta c\right]f_1\nonumber\\
    &+\left[-a\partial_\phi+c\left(-\frac{1}{2}\partial_\theta-\frac{3}{2}\cot\theta\right)-\frac{1}{2}\partial_\theta a\right]f_2 = 0\nonumber,
\end{align}
and
\begin{align}
    (R_{-1})_\phi &= \left[\frac{1}{2}a\partial_\phi+c\left(\partial_\theta +\frac{1}{2}\cot\theta\right)+\frac{1}{2}\partial_\theta c\right]f_1\nonumber\\
    &+\left[a\left(\sin^2\theta\partial_\theta+2\sin\theta\cos\theta\right)-\frac{1}{2}c\partial_\phi+\frac{1}{2}\partial_\theta a\right]f_2 = 0.
\end{align}
\section{Example: $Y^A = z^2\partial_z$}
Here, we work through the example of $Y^A = z^2\partial_z$. We wish to compute $X_{ab} = (1/2)\Omega\cL_\xi\tilde{g}_{ab}$, to verify that the $\Omega^{-1}$ component of $X_{ab}$ is zero and that $X = 0$ at the horizon, as well as compute the charge associated with this value of $Y^A$. First, let us switch coordinates to $(\theta,\phi)$. For this value of $Y^A$, from our expressions for general cases,
\begin{align}
    D_AY^A &= \frac{2z}{1+z\bar{z}} = \sin\theta e^{i\phi},\\
    D^2(D_AY^A) &= -\frac{4z}{1+z\bar{z}} = -2\sin\theta e^{i\phi}.\nonumber
\end{align}
Also,
\begin{equation}
    Y^\theta = -\frac{1}{2}e^{i\phi}(1+\cos\theta), Y^\phi = -\frac{1}{2}ie^{i\phi}\cot(\theta/2).
\end{equation}
Therefore, from (\ref{xi}),
\begin{align}
    \xi_u &= \frac{1}{2}e^{i\phi}u\sin\theta,\nonumber\\
    \xi_\Omega &= \frac{1}{2}e^{i\phi}\Omega(1+u\Omega)\sin\theta + \cO(\Omega^3),\\
    \xi_\theta &= -\frac{1}{4}e^{i\phi}(2+2\cos\theta+2u\Omega\cos\theta-a\cos\theta-ic\csc\theta),\nonumber\\
    \xi_\phi &= -\frac{1}{4}e^{i\phi}i(2+2u\Omega+2\cos\theta+a-ic\cot\theta)\csc\theta.
\end{align}
Note: There are other terms that show up here that have to do with the perturbation $C_{AB}$; however, ultimately they have no relevance to whether or not $X_{ab}$ is well defined, and they do not show up in the constant term of $X = X_{ab}g^{ab}$. First, let's focus on $X_{ab}$. In the Schwarzschild metric, that is, the perturbed metric with $C_{AB} = 0$, we have
\begin{equation}
    X = \frac{1}{2}e^{i\phi}M\Omega^2\begin{pmatrix} (3+u\Omega)\sin\theta & 0 & u\cos\theta & u\sin\theta\\
0 & 0 & 0 & 0\\
u\cos\theta & 0 & 0 & 0\\
u\sin\theta & 0 & 0 & 0\end{pmatrix}.
\end{equation}
Of course, this means that $X_{ab}$ is well defined (no inverse powers of $\Omega$) and that $X = X_{ab}g^{ab} = 0$ at $\cI^+$, since $X_{ab} = 0$ at $\cI^+$. When a nonzero $C_{AB}$ is added, in order to verify that $X_{ab}$ is well defined as well as compute $X = g^{ab}X_{ab}$ at the horizon, we will calculate $X_{ab}$ up to its constant components in $\Omega$. Then, we get, for the nonzero $C_{AB}$ case, up to constant terms in $\Omega$, the only nonzero terms of $X_{ab}$ are
\begin{align}
    X_{\theta\theta} &= \frac{1}{8}e^{i\phi}\csc^2(\theta/2)(2ic-\sin\theta(a(-1+\cos\theta)+\sin\theta\partial_\theta a + i\partial_\phi a)\nonumber\\
    &-u\sin\theta(-1+\cos\theta)\partial_u a),\nonumber\\
X_{\theta\phi} = X_{\phi\theta} &= \frac{1}{4}e^{i\phi}\cot^2(\theta/2)(c-2ia\sin\theta-\sin\theta\partial_\theta c - i\partial_\phi c - u(-1+\cos\theta)\partial_u c),\\
X_{\phi\phi} &= \frac{1}{2}e^{i\phi}\cos^2(\theta/2)(-2ic+\sin\theta(a(-1+\cos\theta)+\sin\theta\partial_\theta a + i\partial_\phi a)\nonumber\\
&+u\sin\theta(-1+\cos\theta)\partial_u a)\nonumber.
\end{align}
Since there are no negative powers of $\Omega$ in this case either, the tensor $X_{ab}$ is well defined. Note that in the Schwarzschild case, we would have gotten zero if we just calculated up to this, so it was actually unnecessary to give the full expression for $X_{ab}$, but that was included for the sake of edification of the reader.

In fact, since we need to just verify that $X = X_{ab}g^{ab} = 0$ on $\cI^+$, and since the constant components of $g^{ab}$ are exactly $g^{u\Omega} = g^{\Omega u} = g^{\theta\theta} = 1$ and $g^{\phi\phi} = \csc^2\theta$, we really only need to compute those corresponding constant components of $X_{ab}$. Therefore, a simple computation tells us that when $C_{AB}$ are nonzero, $X = X_{ab}g^{ab} = 0$ as well.

As for the charge $Q$ corresponding to this choice of $Y^A = z^2\partial_z$, we must calculate
\begin{align}
    D_z(D_A Y^A) &= \frac{2}{(1+z\bar{z})^2} = 2\sin^4(\theta/2),\\
    D_{\bar{z}}(D_A Y^A) &= -\frac{2z^2}{(1+z\bar{z})^2} = -\frac{1}{2}\sin^2\theta e^{2i\phi}.
\end{align}
Therefore,
\begin{align}
    Q &= \frac{1}{8\pi G}\int d^2\Omega \left[Mu(\sin\theta e^{i\phi})-\frac{1}{8}(\sin\theta e^{i\phi})(C_{BC}C^{BC})\right.\nonumber.\\
    &+u U^z\sin^4(\theta/2) -\frac{1}{4}u U^{\bar{z}}\sin^2\theta e^{2i\phi}\nonumber\\
&+3\left(\frac{2}{3}N_z-\frac{1}{16}D_z(C_{BC}C^{BC})-\frac{1}{2}{C_z}^B D^C C_{BC}\right)\cot(\theta/2)^2e^{2i\phi}\\
&+\left.{C^A}_z U_A (\cot(\theta/2)^2e^{2i\phi})+\frac{1}{16}u(\sin\theta e^{i\phi})(C^{CD}\partial_u C_{CD})\right]\nonumber.
\end{align}
Once more, the only terms that could possibly result in divergences are the ones with factors of $\cot(\theta/2)^2$, which are accompanied by factors of $e^{2i\phi}$. We can once more express all of our other variables in terms of spherical harmonics $Y^{\mp 1}_\ell \propto \sin\theta e^{\mp i\phi}$. Once more taking advantage of the fact that $e^{in\phi}$ is periodic in $2\pi$, we obtain that the only surviving integral that we need to consider is
\begin{align}
    \int_0^\pi (\sin\theta d\theta) (\sin^2\theta)\cot^2(\theta/2)&= \frac{8}{3},
\end{align}
which is evidently finite.

\bibliography{references}

@article{Flanagan_2017,
   title={{Conserved charges of the extended Bondi-Metzner-Sachs algebra}},
   volume={95},
   ISSN={2470-0029},
   url={http://dx.doi.org/10.1103/PhysRevD.95.044002},
   DOI={10.1103/physrevd.95.044002},
   number={4},
   journal={Physical Review D},
   publisher={American Physical Society (APS)},
   author={Flanagan, Éanna É. and Nichols, David A.},
   year={2017},
   month=feb }

@article{Pasterski_2019,
   title={Implications of superrotations},
   volume={829},
   ISSN={0370-1573},
   url={http://dx.doi.org/10.1016/j.physrep.2019.09.006},
   DOI={10.1016/j.physrep.2019.09.006},
   journal={Physics Reports},
   publisher={Elsevier BV},
   author={Pasterski, Sabrina},
   year={2019},
   month=oct, pages={1–35} }

@article{Penrose:1962ij,
    author = "Penrose, Roger",
    title = {{Asymptotic properties of fields and space-times}},
    doi = "10.1103/PhysRevLett.10.66",
    journal = "Phys. Rev. Lett.",
    volume = "10",
    pages = "66--68",
    year = "1963"
}

@article{Geroch_1981,
    author = {Geroch, Robert and Winicour, Jeffrey},
    title = {{Linkages in general relativity}},
    journal = {Journal of Mathematical Physics},
    volume = {22},
    number = {4},
    pages = {803-812},
    year = {1981},
    month = apr
}

@article{Barnich_2010b,
   title={{Aspects of the BMS/CFT correspondence}},
   volume={2010},
   ISSN={1029-8479},
   url={http://dx.doi.org/10.1007/JHEP05(2010)062},
   DOI={10.1007/jhep05(2010)062},
   number={5},
   journal={Journal of High Energy Physics},
   publisher={Springer Science and Business Media LLC},
   author={Barnich, Glenn and Troessaert, Cédric},
   year={2010},
   month=may }

@misc{Barnich_2012,
      title={{Supertranslations call for superrotations}}, 
      author={Barnich, Glenn and Troessaert, Cédric},
      year={2012},
      eprint={1102.4632},
      archivePrefix={arXiv},
      primaryClass={gr-qc},
      url={https://arxiv.org/abs/1102.4632}, 
}

@article{Barnich_2011,
   title={{BMS charge algebra}},
   volume={2011},
   ISSN={1029-8479},
   url={http://dx.doi.org/10.1007/JHEP12(2011)105},
   DOI={10.1007/jhep12(2011)105},
   number={12},
   journal={Journal of High Energy Physics},
   publisher={Springer Science and Business Media LLC},
   author={Barnich, Glenn and Troessaert, Cédric},
   year={2011},
   month=dec}

@article{Sachs_1961,
author={Sachs, Rainer},
title={{Gravitational waves in general relativity. VI. The outgoing radiation condition}},
  volume = {264},
  ISSN = {2053-9169},
  url = {http://dx.doi.org/10.1098/rspa.1961.0202},
  DOI = {10.1098/rspa.1961.0202},
  number = {1318},
  journal = {Proceedings of the Royal Society of London. Series A. Mathematical and Physical Sciences},
  publisher = {The Royal Society},
  year = "1961",
  month = nov,
  pages = {309–338}
}

@article{Sachs_1962a,
author={Sachs, Rainer},
title={{Gravitational waves in general relativity. VIII. Waves in Asymptotically Flat Space-Time}},
  volume = {270},
  ISSN = {2053-9169},
  url = {http://dx.doi.org/10.1098/rspa.1962.0206},
  DOI = {10.1098/rspa.1962.0206},
  number = {1340},
  journal = {Proceedings of the Royal Society of London. Series A. Mathematical and Physical Sciences},
  publisher = {The Royal Society},
  year = {1962},
  month = oct,
  pages = {103–126}
}

@article{Sachs_1962b,
author={Sachs, Rainer},
  title = {{Asymptotic symmetries in gravitational theory}},
  volume = {128},
  ISSN = {0031-899X},
  url = {http://dx.doi.org/10.1103/PhysRev.128.2851},
  DOI = {10.1103/physrev.128.2851},
  number = {6},
  journal = {Physical Review},
  publisher = {American Physical Society (APS)},
  year = {1962},
  month = dec,
  pages = {2851–2864}
}

@article{Bondi_1960,
author = {Bondi, Hermann},
  title = {Gravitational Waves in General Relativity},
  volume = {186},
  ISSN = {1476-4687},
  url = {http://dx.doi.org/10.1038/186535a0},
  DOI = {10.1038/186535a0},
  number = {4724},
  journal = {Nature},
  publisher = {Springer Science and Business Media LLC},
  year = {1960},
  month = may,
  pages = {535–535}
}

@article{Bondi_1962,
author = {Bondi, Hermann and Van der Burg, M. G. J. and Metzner, A. W. K.},
year = {1962},
month = {aug},
title = {{Gravitational waves in general relativity, VII. Waves from axi-symmetric isolated system}},
journal = {Proceedings of the Royal Society of London. Series A. Mathematical and Physical Sciences},
publisher = {The Royal Society},
DOI = {10.1098/rspa.1962.0161},
volume = {269},
issue = {1336},
pages = {21-52}
}

@misc{Banks_2003,
      title={{A Critique of Pure String Theory: Heterodox Opinions of Diverse Dimensions}}, 
      author={Banks, T.},
      year={2003},
      eprint={hep-th/0306074},
      archivePrefix={arXiv},
      primaryClass={hep-th},
      url={https://arxiv.org/abs/hep-th/0306074}, 
}

@article{Wald_2000,
   title={General definition of “conserved quantities” in general relativity and other theories of gravity},
   volume={61},
   ISSN={1089-4918},
   url={http://dx.doi.org/10.1103/PhysRevD.61.084027},
   DOI={10.1103/physrevd.61.084027},
   number={8},
   journal={Physical Review D},
   publisher={American Physical Society (APS)},
   author={Wald, Robert M. and Zoupas, Andreas},
   year={2000},
   month=mar }

@techreport{Crnkovic_1986,
      author        = {Crnkovic, C and Witten, Edward},
      title         = {Covariant description of canonical formalism in
                       geometrical theories},
      institution   = "Princeton Univ.",
      address       = "Princeton, NJ",
      year          = "1986",
      url           = "https://cds.cern.ch/record/172498",
}

@incollection{Capone_2018,
    author = "Capone, Federico",
    editor = {Cacciatori, Sergio and G\"uneysu, Batu and Pigola, Stefano},
    title = "{BMS Symmetries and Holography: An Introductory Overview}",
    booktitle = "{Einstein Equations: Physical and Mathematical Aspects of General Relativity. Domoschool 2018}",
    publisher = {Birkhäuser, Cham},
    doi = "10.1007/978-3-030-18061-4_6",
    pages = "197--225",
    month = "5",
    year = "2018"
}

@article{Campiglia_2014,
   title={Asymptotic symmetries and subleading soft graviton theorem},
   volume={90},
   ISSN={1550-2368},
   url={http://dx.doi.org/10.1103/PhysRevD.90.124028},
   DOI={10.1103/physrevd.90.124028},
   number={12},
   journal={Physical Review D},
   publisher={American Physical Society (APS)},
   author={Campiglia, Miguel and Laddha, Alok},
   year={2014},
   month=dec }

@article{Campiglia_2015,
   title={{New symmetries for the gravitational S-matrix}},
   volume={2015},
   ISSN={1029-8479},
   url={http://dx.doi.org/10.1007/JHEP04(2015)076},
   DOI={10.1007/jhep04(2015)076},
   number={4},
   journal={Journal of High Energy Physics},
   publisher={Springer Science and Business Media LLC},
   author={Campiglia, Miguel and Laddha, Alok},
   year={2015},
   month=apr }

@article{Strominger_2017,
   title={{Superrotations and black hole pair creation}},
   volume={34},
   ISSN={1361-6382},
   url={http://dx.doi.org/10.1088/1361-6382/aa5b5f},
   DOI={10.1088/1361-6382/aa5b5f},
   number={6},
   journal={Classical and Quantum Gravity},
   publisher={IOP Publishing},
   author={Strominger, Andrew and Zhiboedov, Alexander},
   year={2017},
   month=feb, pages={064002} }

@misc{Cachazo_2014,
      title={{Evidence for a New Soft Graviton Theorem}}, 
      author={Freddy Cachazo and Andrew Strominger},
      year={2014},
      eprint={1404.4091},
      archivePrefix={arXiv},
      primaryClass={hep-th},
      url={https://arxiv.org/abs/1404.4091}, 
}

@article{Compere_2018,
   title={{Superboost transitions, refraction memory and super-Lorentz charge algebra}},
   volume={2018},
   ISSN={1029-8479},
   url={http://dx.doi.org/10.1007/JHEP11(2018)200},
   DOI={10.1007/jhep11(2018)200},
   number={11},
   journal={Journal of High Energy Physics},
   publisher={Springer Science and Business Media LLC},
   author={Compère, Geoffrey and Fiorucci, Adrien and Ruzziconi, Romain},
   year={2018},
   month=nov }

@article{Chandrasekaran_2018,
  title = {{Symmetries and charges of general relativity at null boundaries}},
  volume = {2018},
  ISSN = {1029-8479},
  url = {http://dx.doi.org/10.1007/JHEP11(2018)125},
  DOI = {10.1007/jhep11(2018)125},
  number = {11},
  journal = {Journal of High Energy Physics},
  publisher = {Springer Science and Business Media LLC},
  author = {Chandrasekaran,  Venkatesa and Flanagan,  Éanna É. and Prabhu,  Kartik},
  year = {2018},
  month = nov 
}

@article{Komar_1958,
    author = "Komar, Arthur",
    title = "{Covariant conservation laws in general relativity}",
    doi = "10.1103/PhysRev.113.934",
    journal = "Phys. Rev.",
    volume = "113",
    pages = "934--936",
    year = "1959"
}

@article{Compere_2020,
   title={{The $\Lambda-\text{BMS}_4$ charge algebra}},
   volume={2020},
   ISSN={1029-8479},
   url={http://dx.doi.org/10.1007/JHEP10(2020)205},
   DOI={10.1007/jhep10(2020)205},
   number={10},
   journal={Journal of High Energy Physics},
   publisher={Springer Science and Business Media LLC},
   author={Compère, Geoffrey and Fiorucci, Adrien and Ruzziconi, Romain},
   year={2020},
   month=oct }

@article{Freidel_2021,
   title={{The Weyl BMS group and Einstein’s equations}},
   volume={2021},
   ISSN={1029-8479},
   url={http://dx.doi.org/10.1007/JHEP07(2021)170},
   DOI={10.1007/jhep07(2021)170},
   number={7},
   journal={Journal of High Energy Physics},
   publisher={Springer Science and Business Media LLC},
   author={Freidel, Laurent and Oliveri, Roberto and Pranzetti, Daniele and Speziale, Simone},
   year={2021},
   month=jul }

@misc{RignonBret_2025,
      title={General covariance and boundary symmetry algebras}, 
      author={Antoine Rignon-Bret and Simone Speziale},
      year={2025},
      eprint={2403.00730},
      archivePrefix={arXiv},
      primaryClass={hep-th},
      url={https://arxiv.org/abs/2403.00730}, 
}

@misc{RignonBret_2024,
      title={{Center-less BMS charge algebra}}, 
      author={Antoine Rignon-Bret and Simone Speziale},
      year={2024},
      eprint={2405.01526},
      archivePrefix={arXiv},
      primaryClass={hep-th},
      url={https://arxiv.org/abs/2405.01526}, 
}

@misc{Fiorucci_2021,
      title={{Leaky covariant phase spaces: Theory and application to $\Lambda$-BMS symmetry}}, 
      author={Adrien Fiorucci},
      year={2021},
      eprint={2112.07666},
      archivePrefix={arXiv},
      primaryClass={hep-th},
      url={https://arxiv.org/abs/2112.07666}, 
}

@article{Hawking_2016,
    author = "Hawking, Stephen W. and Perry, Malcolm J. and Strominger, Andrew",
    title = "{Superrotation Charge and Supertranslation Hair on Black Holes}",
    eprint = "1611.09175",
    archivePrefix = "arXiv",
    primaryClass = "hep-th",
    doi = "10.1007/JHEP05(2017)161",
    journal = "JHEP",
    volume = "05",
    pages = "161",
    year = "2017"
}

@book{Strominger_2017a,
    author = "Strominger, Andrew",
    title = "{Lectures on the Infrared Structure of Gravity and Gauge Theory}",
    eprint = "1703.05448",
    archivePrefix = "arXiv",
    primaryClass = "hep-th",
    isbn = "978-0-691-17973-5",
    publisher = "Princeton University Press",
    year = "2018"
}

@article{Chen:2022fbu,
    author = "Chen, Po-Ning and Paraizo, Daniel E. and Wald, Robert M. and Wang, Mu-Tao and Wang, Ye-Kai and Yau, Shing-Tung",
    title = "{Cross-section continuity of definitions of angular momentum}",
    eprint = "2207.04590",
    archivePrefix = "arXiv",
    primaryClass = "gr-qc",
    doi = "10.1088/1361-6382/acaa82",
    journal = "Class. Quant. Grav.",
    volume = "40",
    number = "2",
    pages = "025007",
    year = "2023"
}

@article{Chen:2013kza,
    author = "Chen, Po-Ning and Wang, Mu-Tao and Yau, Shing-Tung",
    title = "{Conserved quantities in general relativity: from the quasi-local level to spatial infinity}",
    eprint = "1312.0985",
    archivePrefix = "arXiv",
    primaryClass = "math.DG",
    doi = "10.1007/s00220-015-2381-1",
    journal = "Commun. Math. Phys.",
    volume = "338",
    number = "1",
    pages = "31--80",
    year = "2015"
}

@article{Chen:2013lza,
    author = "Chen, Po-Ning and Wang, Mu-Tao and Yau, Shing-Tung",
    title = "{Quasilocal angular momentum and center of mass in general relativity}",
    eprint = "1312.0990",
    archivePrefix = "arXiv",
    primaryClass = "math.DG",
    doi = "10.4310/ATMP.2016.v20.n4.a1",
    journal = "Adv. Theor. Math. Phys.",
    volume = "20",
    pages = "671--682",
    year = "2016"
}

@article{Flanagan_2020,
   title={Extensions of the asymptotic symmetry algebra of general relativity},
   volume={2020},
   ISSN={1029-8479},
   url={http://dx.doi.org/10.1007/JHEP01(2020)002},
   DOI={10.1007/jhep01(2020)002},
   number={1},
   journal={Journal of High Energy Physics},
   publisher={Springer Science and Business Media LLC},
   author={Flanagan, Éanna É. and Prabhu, Kartik and Shehzad, Ibrahim},
   year={2020},
   month=jan }

\end{document}